\documentclass[aps,prd,amsmath,amssymb,preprintnumbers,nofootinbib,11pt]{revtex4-2}

\usepackage[T1]{fontenc}
\usepackage{graphicx}
\usepackage{bm}
\usepackage{hyperref}
\usepackage{color}
\usepackage{mathtools}
\usepackage{slashed}
\usepackage{physics}
\usepackage{float}
\usepackage{caption}
\usepackage{amsmath}
\allowdisplaybreaks
\newcommand{\Lterm}{\frac{\Lambda^4 m^2 Q^2}{\Lambda_c^{12} a^6}}  
\newcommand{\Hubble}{H_0}
\newcommand{\mQ}{\sqrt{\frac{Q}{m}}}

\allowdisplaybreaks

\begin{document}

\title{Relativistic Corrections and Structure Formation\\in Dark Matter Superfluidity}

\author{Seturumane Tema}
\email{tmxset001@myuct.ac.za}
\affiliation{Department of Mathematics and Applied Mathematics, University of Cape Town, Rondebosch 7701, South Africa}

\begin{abstract}
The theory of dark matter superfluidity has emerged as a compelling framework, in which the dynamics are governed by a non-relativistic $P(X)$ superfluid Lagrangian that naturally leads to Modified Newtonian Dynamics (MOND-like behavior) when coupled to baryons at galactic scales. Notably, at cosmological scales, this effective description reproduces the standard $\Lambda$ Cold Dark Matter ($\Lambda$CDM) model at the background level, suggesting that cold dark matter may undergo Bose--Einstein condensation at galactic scales. In this work, we extend the non-relativistic formulation by incorporating relativistic corrections within the Friedmann--Lemaître--Robertson--Walker (FLRW) spacetime. We further perform a linear perturbation analysis in this relativistic setting to investigate the evolution of matter density fluctuations. Our results clarify the viability of the superfluid dark matter scenario in explaining large-scale structure formation and identify the parameter regimes in which it remains consistent with current cosmological observations.
\end{abstract}

\maketitle
\tableofcontents
\section{Introduction}

Within the framework of the standard cosmological model, the $\Lambda$CDM paradigm posits that dark matter constitutes approximately 25\% of the total energy budget of the Universe. Despite its central role in explaining a wide array of cosmological observations, the fundamental nature of dark matter remains one of the most profound open questions in contemporary physics \cite{peebles2015dark, ellis2012relativistic}. On large scales, dark matter is effectively described as a pressureless perfect fluid with an equation of state parameter $w \approx 0$, a treatment that successfully reproduces the observed matter power spectrum and aligns with precision measurements of the cosmic microwave background and large-scale structure.
\\
\\
On galactic scales, however, the $\Lambda$CDM model encounters notable challenges. In particular, it struggles to account for the observed flatness of galactic rotation curves and the empirical success of the Baryonic Tully–Fisher relation \cite{rubin1970rotation, freeman2001dark, mcgaugh2000baryonic, mcgaugh2016radial}. While these phenomena are broadly compatible with the presence of extended dark matter halos, the tightness of the observed correlations and the detailed shapes of rotation curves especially in low surface brightness and dwarf galaxies are not easily derived from first principles within $\Lambda$CDM and typically require fine-tuned baryonic feedback mechanisms in numerical simulations \cite{governato2010bulgeless, dicintio2014core, trujillo2016cold}. These shortcomings have led to growing interest in alternative models that aim to address such galactic-scale anomalies more directly \cite{famaey2012modified, milgrom2020mondreview}.
\\
\\
In contrast, Modified Newtonian Dynamics (MOND) \cite{milgrom2008mond, bekenstein1984does} has been remarkably successful in explaining many of these galactic phenomena without invoking dark matter, though it lacks a consistent relativistic and cosmological extension. A promising attempt to reconcile these seemingly divergent approaches involves the proposal that dark matter forms a Bose–Einstein condensate within galactic halos, transitioning into a superfluid phase \cite{r1, berezhiani2018phenomenological}. In this regime, dark matter exhibits collective behavior governed by a low-energy effective field theory of superfluidity, in which phonon excitations mediate a MOND-like force between baryons \cite{khoury2015dark, berezhiani2015theory, berezhiani2015phenomenology}. This paradigm naturally interpolates between MOND phenomenology on galactic scales and $\Lambda$CDM dynamics on cosmological scales, unifying them as emergent limits of a single underlying theory.
\\
\\
The present work explores the relativistic formulation of this superfluid dark matter scenario in the context of cosmology. In Section~II, we review the non-relativistic effective field theory that underpins the superfluid description. Section~III extends this framework to include relativistic corrections in a Friedmann–Lemaître–Robertson–Walker (FLRW) background. Sections~IV and V present a detailed linear perturbation analysis of the model to assess the evolution of structure at late times. We conclude in Section~VI with a discussion of our results and their implications for the viability of the superfluid dark matter paradigm.

\section{Non-Relativistic Superfluid Framework}

The dynamics of a superfluid dark matter \cite{r1, berezhiani2018phenomenological, mistele2022galactic, khoury2022dark} are governed by a non-relativistic effective field theory, whose Lagrangian density is given by
\begin{equation}\label{rellagr}
\mathcal{L} = \frac{2\Lambda (2m)^{3/2}}{3} \left( \dot{\theta} - m\Phi - \frac{(\nabla \theta)^2}{2m} \right)^{3/2},
\end{equation}
where $\theta$ denotes the Goldstone boson (phonon) field of the superfluid, $m$ is the phonon mass, and $\Phi$ is the Newtonian gravitational potential. The Lagrangian \eqref{rellagr} is a specific instance of the general effective field theory of superfluidity, which takes the functional form
\begin{equation}
\mathcal{L} = P(X), \quad \text{where} \quad X = \dot{\theta} - m\Phi - \frac{(\nabla \theta)^2}{2m},
\end{equation}
with $P(X)$ interpreted as the superfluid pressure. To reproduce the correct MOND-like force law in the presence of baryons, the \(P(X)\) function must scale as $P(X) \propto X^{3/2}$ \cite{r1}, as other powers of \(X\) fail to reproduce the desired phenomenology. A convenient decomposition of the phonon field is
\begin{equation}\label{aust}
\theta = \mu t + \psi,
\end{equation}
where $\mu$ represents the chemical potential, and $\psi$ encodes the dynamical phonon excitations. Substituting \eqref{aust} into \eqref{rellagr} lead us to
\begin{equation}\label{mar}
\mathcal{L} = \frac{2\Lambda (2m)^{3/2}}{3} \left( \mu + \dot{\psi} - m\Phi - \frac{(\nabla \psi)^2}{2m} \right)^{3/2}.
\end{equation}
In the limit where both the gravitational potential and phonon fluctuations vanish, equation \eqref{mar} becomes
\begin{equation}
\mathcal{L} = P(\mu) = \frac{2\Lambda (2m)^{3/2}}{3} \mu^{3/2},
\end{equation}
which corresponds to the grand canonical equation of state of the superfluid dark matter, consistent with results from condensed matter systems \cite{son2002low, greiter1989hydrodynamic}. Thermodynamic quantities such as the number density \(n\) of the superfluid can be derived from this relation.
To enable a MOND-like interaction between the superfluid dark matter and baryons, the phonon field couples to the baryonic mass density $\rho_b$ via the interaction term \(\alpha \frac{\theta \rho_b}{M_{\text{Pl}}}\) in the manner
\begin{equation}\label{zoo}
\mathcal{L} = \frac{2\Lambda (2m)^{3/2}}{3} \left( \dot{\theta} - m\Phi - \frac{(\nabla \theta)^2}{2m} \right)^{3/2} - \alpha \frac{\theta \rho_b}{M_{\text{Pl}}},
\end{equation}
where $\alpha$ is a dimensionless coupling constant, and $M_{\text{Pl}}$ denotes the Planck mass. It was demonstrated in \cite{r1} that this interaction gives rise to the MOND acceleration law \cite{milgrom1983modification, famaey2012modified} in the static, spherically symmetric limit, provided that $X$ is negative and
\begin{equation}\label{osmo}
\alpha = 0.86 \left( \frac{\Lambda}{\mathrm{meV}} \right)^{-3/2}.
\end{equation}
This confirms that the phonon excitations of the dark matter superfluid mediate a fifth force consistent with MOND-like dynamics at galactic scales, thereby unifying both approaches within a single coherent framework. At cosmological scales, the dynamics are embedded in an expanding Friedmann–Lemaître–Robertson–Walker (FLRW) background. The equation of motion derived from the Lagrangian \eqref{zoo} at cosmological scales reads
\begin{equation}
\frac{d}{dt} \left[ (2m)^{3/2} a^3 \dot{\theta}^{1/2} \right] = - \frac{\alpha}{M_{\text{Pl}}} a^3 \rho_b,
\end{equation}
where $a(t)$ is the cosmological scale factor. Upon integration, the energy density of superfluid dark matter becomes
\begin{equation}\label{zss}
\rho = - \frac{\alpha \Lambda_0}{M_{\text{Pl}}} m \rho_b t + \frac{m \Lambda_0 C}{a^3},
\end{equation}
where $C$ is an integration constant, $\Lambda=\Lambda_0$ at cosmological scales and $\rho=mn=m\Lambda_0\left(2m\right)^{\frac{3}{2}}\dot{\theta}^{\frac{1}{2}}$ \cite{r1}. The second term in \eqref{zss} dominates when the coupling $\alpha$ is sufficiently small, in which case the superfluid dark matter behaves as a pressureless dust. This happens when:
\begin{equation}
\alpha \ll 2.4 \times 10^{-4} \left( \frac{m}{\mathrm{eV}} \right)^2,
\end{equation}
a condition that is more stringent than the galactic-scale constraint \eqref{osmo}. In summary, this framework suggests that dark matter undergoes a phase transition into a superfluid state within galactic halos, thereby giving rise to MOND-like behavior via phonon-mediated interactions with baryons. At larger scales, dark matter retains its standard cold, pressureless character, ensuring consistency with cosmological observations. The detailed microphysical mechanism underlying the onset of the phase transition, including its dependence on environmental conditions, remains an open area of investigation. Finally, it is worth noting that the non-relativistic Lagrangian \eqref{rellagr} can be derived as the weak-field limit of the full relativistic theory
\begin{equation}\label{asw}
\mathcal{L} = -\frac{1}{2} \mathcal{L}_{\text{free}} - \frac{\Lambda^4}{(\Lambda_c^2+ |\phi|^2)^6} \mathcal{L}_{\text{free}}^3,
\end{equation}
where
\begin{equation}
\mathcal{L}_{\text{free}} = -|\partial_\mu \phi|^2 + m^2 |\phi|^2,
\end{equation}
and $\phi$ is a complex scalar field \cite{r1}. The forthcoming sections are devoted to analyzing the cosmological implications of the relativistic theory \eqref{asw} in an FLRW background, with particular emphasis on its impact on the growth of structure and consistency with observations.
\section{Relativistic background}

When the relativistic Lagrangian \eqref{asw} is varied with respect to $\phi^*$, in the regime $\Lambda_c \gg |\phi|^2$ (opposite to the galactic scale limit considered in \cite{r1}), it results in the equation of motion:
\begin{multline}\label{lasr}
\frac{1}{2} \Box \phi + \frac{3\Lambda^4}{\Lambda_c^{12}} \Box \phi \left( g^{\mu\nu} \partial_\mu \phi \partial_\nu \phi^* + m^2 \phi \phi^* \right)^2 \\
= -\frac{1}{2} m^2 \phi - \frac{3\Lambda^4 m^2 \phi}{\Lambda_c^{12}} \left( g^{\mu\nu} \partial_\mu \phi \partial_\nu \phi^* + m^2 \phi \phi^* \right)^2 + \frac{6\Lambda^4 \phi}{\Lambda_c^{14}} \left( g^{\mu\nu} \partial_\mu \phi \partial_\nu \phi^* + m^2 \phi \phi^* \right)^3.
\end{multline}

Expressing the complex scalar field \(\phi\) in terms of its modulus \(\bar{\rho}\) and phase \(\bar{\lambda}\),
\[
\phi(t) = \bar{\rho}(t) e^{i \bar{\lambda}(t)},
\]
the imaginary part of \eqref{lasr} in a Friedmann--Lemaître--Robertson--Walker (FLRW) spacetime becomes
\begin{align}\label{zsr}
&\left( 2 \dot{\bar{\rho}} \dot{\bar{\lambda}} + \bar{\rho} \ddot{\bar{\lambda}} + 3 H \bar{\rho} \dot{\bar{\lambda}} \right) \times 
\left[ \frac{1}{2} + \frac{3\Lambda^4}{\Lambda_c^{12}} \left( \dot{\bar{\rho}}^2 + \bar{\rho}^2 \dot{\bar{\lambda}}^2 + m^2 \bar{\rho}^2 \right) \right] = 0,
\end{align}
where $H = \frac{\dot{a}}{a}$ is the Hubble parameter and the overdots denote derivatives with respect to the cosmic time $t$. Equation \eqref{zsr} holds if either
\begin{equation}\label{ezi}
2 \dot{\bar{\rho}} \dot{\bar{\lambda}} + \bar{\rho} \ddot{\bar{\lambda}} + 3 H \bar{\rho} \dot{\bar{\lambda}} = 0
\end{equation}
or
\begin{equation}\label{aze}
\frac{1}{2} + \frac{3\Lambda^4}{\Lambda_c^{12}} \left( \dot{\bar{\rho}}^2 + \bar{\rho}^2 \dot{\bar{\lambda}}^2 + m^2 \bar{\rho}^2 \right)^2 = 0.
\end{equation}
Equation \eqref{aze} does not admit any real solutions. The expression on the left-hand side consists of a positive constant, \(\frac{1}{2}\), added to a strictly non-negative term: a squared real-valued expression scaled by a positive function. This sum is therefore strictly greater than zero for all real values of \(\bar{\rho}\), \(\dot{\bar{\rho}}\), and \(\dot{\bar{\lambda}}\), and can never vanish. Thus, Equation \eqref{aze} is incompatible with physically admissible field configurations and is excluded from further analysis. We thus proceed with the alternative condition given by Equation \eqref{ezi}, which leads to consistent and solvable dynamics. Equation \eqref{ezi} can be rearranged as
\begin{equation}\label{ozo}
\frac{d\bar{\rho}}{\bar{\rho}} + \frac{1}{2} \frac{d \dot{\bar{\lambda}}}{\dot{\bar{\lambda}}} + \frac{3}{2} \frac{d a}{a} = 0.
\end{equation}
Integration gives
\begin{equation}
\ln \left( \bar{\rho}^2 \dot{\bar{\lambda}} a^3 \right) = C,
\end{equation}
with $C$ an integration constant. Setting the phase velocity $\dot{\bar{\lambda}} = m$ as in \cite{r1}, we obtain
\begin{equation}
\bar{\rho} = \sqrt{\frac{Q}{m}}\, a^{-\frac{3}{2}}
\end{equation}
where $Q = e^{C}$. The zero-zero component of the energy-momentum tensor derived from \eqref{asw} corresponds to the energy density
\begin{equation}\label{zqw}
\bar{\mu} = \frac{1}{2} \dot{\bar{\rho}}^2 + m^2 \bar{\rho}^2 + \frac{\Lambda^4}{\Lambda_c^{12}} \bar{\rho}^2 m^2 \dot{\bar{\rho}}^4 + \frac{5 \Lambda^4 \dot{\bar{\rho}}^6}{6 \Lambda_c^{12}}.
\end{equation}
The pressure of the scalar field is given by
\begin{equation}
\bar{p} = \frac{1}{2} \dot{\bar{\rho}}^2 + \frac{\Lambda^4 \dot{\bar{\rho}}^6}{6 \Lambda_c^{12}}.
\end{equation}

For the complex scalar field to mimic dust, the term
\[
m^2 \bar{\rho}^2 = m Q a^{-3}
\]
must dominate all other terms in \eqref{zqw}, leading to the constraint
\begin{equation}\label{afg}
\left( \frac{m}{H_0} \right)^2 \gg \frac{9}{8} (1 + z_{\text{eq}})^3 \simeq 10^9,
\end{equation}
where $z_{\text{eq}}$ is the redshift at matter-radiation equality and $H_0$ is the Hubble parameter today. In the next section, we investigate structure formation based on \eqref{asw} under the above constraint \eqref{afg}.

\section{Linear Perturbation Analysis in the Relativistic Regime}
To examine the evolution of linear perturbations within the relativistic regime of the superfluid dark matter model, we consider first-order perturbations of the complex scalar field equation of motion~\eqref{lasr} in the Newtonian gauge. The scalar field is expanded as
\begin{equation}
\phi(t, \vec{x}) = \bar{\phi}(t) + \delta\phi(t, \vec{x}),
\end{equation}
where \(\bar{\phi}(t)\) is the homogeneous complex background field and \(\delta\phi(t, \vec{x})\) is a complex-valued first order perturbation of \(\bar{\phi} (t)\). This reduces (\ref{lasr}) to
\begin{widetext}
\begin{multline}\label{zrr}
\frac{1}{2} \bar{g}^{\mu\nu} \nabla_{\mu} \nabla_{\nu} \delta \phi 
+ \frac{1}{2} \delta g^{\mu\nu} \nabla_{\mu} \nabla_{\nu} \bar{\phi}
+ \frac{\Lambda^4}{2 \Lambda_c^{12}} \bar{g}^{\mu\nu} \nabla_{\mu} \nabla_{\nu} \delta \phi 
\left( \bar{g}^{\mu\nu} \partial_{\mu} \bar{\phi} \partial_{\nu} \bar{\phi} + m^2 \bar{\phi} \bar{\phi}^* \right)^2 \\
+ \frac{\Lambda^4}{2 \Lambda_c^{12}} \bar{g}^{\mu\nu} \nabla_{\mu} \nabla_{\nu} \bar{\phi} 
\delta \left( g^{\mu\nu} \partial_{\mu} \phi \partial_{\nu} \phi + m^2 \phi \phi^* \right)^2
+ \frac{\Lambda^4}{2 \Lambda_c^{12}} \delta g^{\mu\nu} \nabla_{\mu} \nabla_{\nu} \bar{\phi} 
\left( \bar{g}^{\mu\nu} \partial_{\mu} \bar{\phi} \partial_{\nu} \bar{\phi} + m^2 \bar{\phi} \bar{\phi}^* \right)^2 \\
= -\frac{1}{2} m^2 \delta \phi 
- \frac{\Lambda^4}{2 \Lambda_c^{12}} m^2 \delta \phi 
\left( \bar{g}^{\mu\nu} \partial_{\mu} \bar{\phi} \partial_{\nu} \bar{\phi} + m^2 \bar{\phi} \bar{\phi}^* \right)^2
- \frac{\Lambda^4}{2 \Lambda_c^{12}} m^2 \bar{\phi} 
\delta \left( g^{\mu\nu} \partial_{\mu} \phi \partial_{\nu} \phi + m^2 \phi \phi^* \right)^2 \\
+ \frac{\Lambda^4}{2 \Lambda_c^{12}} \bar{\phi} 
\delta \left( g^{\mu\nu} \partial_{\mu} \phi \partial_{\nu} \phi + m^2 \phi \phi^* \right)^3
+ \frac{\Lambda^4}{2 \Lambda_c^{12}} \delta \phi 
\left( \bar{g}^{\mu\nu} \partial_{\mu} \bar{\phi} \partial_{\nu} \bar{\phi} + m^2 \bar{\phi} \bar{\phi}^* \right)^3.
\end{multline}
\end{widetext}
In the Newtonian gauge, the perturbed metric takes the form
\begin{equation}\label{pser}
ds^2 = -(1 + 2\Psi)\,dt^2 + a^2(t)(1 - 2\Phi)\,d\vec{x}^2.
\end{equation}
We assume that the anisotropic stress vanishes, as is typical for minimally coupled scalar fields and perfect fluids, allowing us to set \(\Psi = \Phi\). The decomposition of the scalar field into its modulus and phase, \(\bar{\phi} = \bar{\rho} e^{i \bar{\lambda}}\), leads to two coupled equations governing the perturbations of the amplitude and phase:
\vspace{6pt}

\noindent\textbf{Real part equation:}

\begin{align}\label{ovlt}
& \left[1 + \Lterm \right] \frac{d^2 \delta \rho}{d a^2} 
+ \frac{5}{2 a} \biggl[1 + \Lterm - \frac{36 \Lambda^4 m^2 Q^2}{5 \Lambda_c^{12} a^9} \nonumber \\
& \quad - \frac{36 \Lambda^4 m^2 Q^3}{5 \Lambda_c^{14} a^9} + \frac{12 \Lambda^4 m^2 Q^2}{5 \Lambda_c^{14} a^6} 
+ \frac{24 \Lambda^4 m^2 Q^2}{5 \Lambda_c^{12} a^6} \biggr] \frac{d \delta \rho}{d a} \nonumber \\
& + \biggl[ - \frac{2 m \mQ}{\Hubble a} \left(1 + \Lterm \right) 
- \frac{12 \Lambda^4 Q^2 \Hubble m}{\Lambda_c^{12} a^{10}} \mQ 
+ \frac{12 \Lambda^4 m^3 Q^2}{\Hubble \Lambda_c^{12} a^7} \biggr] \frac{d \delta \lambda}{d a} \nonumber \\
& - \frac{a}{\Hubble^2} \biggl[-k^2 + 2 m^2 + \Lterm (-k^2 + m^2) 
- \frac{\Lambda^4 m^4 Q^2}{\Lambda_c^{12} a^6} + \frac{6 \Lambda^4 m^3 Q^3}{\Lambda_c^{12} a^9} \biggr] \delta \rho \nonumber \\
& - \Phi \mQ \biggl[ \left( \frac{3}{a^{7/2}} - \frac{2}{a^{1/2}} \left(\frac{m}{\Hubble}\right)^2 \right) \left( 1 - \frac{10 \Lambda^4 m^2 Q^2}{\Lambda_c^{12} a^6} \right) \nonumber \\
& \quad + \frac{4 \Lambda^4 m^2 Q^2}{\Lambda_c^{12} a^{13/2}} \left( \frac{m}{\Hubble} \right)^2
- \frac{12 \Lambda^4 m^3 Q^3}{\Lambda_c^{14} \Hubble^2 a^{19/2}} \biggr] = 0.
\end{align}

\vspace{6pt}

\noindent\textbf{Imaginary part equation:}

\begin{align}\label{awr}
& \left[1 + \Lterm \right] \frac{d^2 \delta \lambda}{d a^2} 
- \frac{1}{2 a} \left[1 + \frac{2 \Lambda^4 m^2 Q^2}{\Lambda_c^{12} a^5} - \frac{12 \Lambda^4 m^2 Q^2}{\Lambda_c^{12} a^6} \right] \frac{d \delta \lambda}{d a} \nonumber \\
& - \frac{a}{\Hubble^2} \left[-k^2 + 2 m^2 + \Lterm (-k^2 + m^2) + \frac{\Lambda^4 m^4 Q^2}{\Lambda_c^{12} a^4} \right] \delta \lambda \nonumber \\
& + \left[ \frac{2 m}{\Hubble \mQ} a^2 \left(1 + \Lterm \right) 
+ \frac{18 \Lambda^4 Q \Hubble m^2}{\Lambda_c^{12} a^8} \mQ \right] \frac{d \delta \rho}{d a} \nonumber \\
& + \left[ \frac{3 m}{\Hubble \mQ} a \left( 1 + \Lterm \right) \right] \delta \rho
+ 6 \Phi \frac{m}{\Hubble} a^{-12} \left(1 + \Lterm \right) = 0.
\end{align}

\vspace{6pt}

The gravitational potential \(\Phi\) in the weak-field limit is determined by Poisson's equation:

\begin{equation}\label{poisson}
\Phi = - \frac{4 \pi G}{k^2} a^2 \delta \mu,
\end{equation}

where the perturbed energy density \(\delta \mu\) is obtained from the perturbation of the energy-momentum tensor \(\delta T_{\mu\nu}\) of the scalar field:

\begin{multline}\label{eq:tmunu}
\delta T_{\mu\nu} = \delta \left( \partial_{\mu} \phi \partial_{\nu} \phi^* \right)
+ \frac{\Lambda^4}{\Lambda_c^{12}} \delta \left( \partial_{\mu} \phi \partial_{\nu} \phi^* \right) \left( \bar{g}^{\alpha\beta} \partial_{\alpha} \bar{\phi} \partial_{\beta} \bar{\phi}^* + m^2 \bar{\phi} \bar{\phi}^* \right)^2 \\
+ \frac{\Lambda^4}{\Lambda_c^{12}} \partial_{\mu} \bar{\phi} \partial_{\nu} \bar{\phi}^* 
\delta \left( g^{\alpha\beta} \partial_{\alpha} \phi \partial_{\beta} \phi^* + m^2 \phi \phi^* \right)^2
- \frac{1}{2} \delta g_{\mu\nu} \left( \bar{g}^{\alpha\beta} \partial_{\alpha} \bar{\phi} \partial_{\beta} \bar{\phi}^* + m^2 \bar{\phi} \bar{\phi}^* \right) \\
- \frac{1}{2} \bar{g}_{\mu\nu} \delta \left( g^{\alpha\beta} \partial_{\alpha} \phi \partial_{\beta} \phi^* + m^2 \phi \phi^* \right)
- \frac{\Lambda^4}{6 \Lambda_c^{12}} \delta g_{\mu\nu} \left( \bar{g}^{\alpha\beta} \partial_{\alpha} \bar{\phi} \partial_{\beta} \bar{\phi}^* + m^2 \bar{\phi} \bar{\phi}^* \right)^3 \\
- \frac{\Lambda^4}{\Lambda_c^{12}} \bar{g}_{\mu\nu} \delta \left( g^{\alpha\beta} \partial_{\alpha} \phi \partial_{\beta} \phi^* + m^2 \phi \phi^* \right)^3.
\end{multline}

The zero-zero component explicitly reads:

\begin{align}
\delta T_{00} &= \frac{3 Q \Hubble}{a^{3/2}} \frac{d \delta \lambda}{d a}
- \frac{9 \Hubble^2}{2 a^{3/2}} \mQ \frac{d \delta \rho}{d a}
+ 2 m^2 \mQ a^{1/2} \delta \rho
+ \frac{3 \Phi m Q}{a} \nonumber \\
& + \frac{\Lambda^4 Q^2 m^2}{\Lambda_c^{12} a^6} \left( - \frac{3 \Hubble^2}{a^{7/2}} \mQ \frac{d \delta \rho}{d a} + \frac{2 Q \Hubble}{a^{7/2}} \frac{d \delta \lambda}{d a} + \frac{2 m^2}{a^{1/2}} \mQ \delta \rho \right) \nonumber \\
& + \frac{2 \Lambda^4 Q m}{a \Lambda_c^{12}} \left( \frac{3 \Hubble^2}{a^{7/2}} \mQ \frac{d \delta \rho}{d a} - \frac{2 Q \Hubble}{a^{7/2}} \frac{d \delta \lambda}{d a} + \frac{2 \Phi Q m}{a^3} \right) \nonumber \\
& + \frac{\Lambda^4 Q^3 m^3}{3 \Lambda_c^{12} a^7} \Phi 
+ \frac{3 \Lambda^4 m^2 Q^2}{\Lambda_c^{12} a^6} \left( \frac{3 \Hubble^2}{a^{7/2}} \mQ \frac{d \delta \rho}{d a} - \frac{2 Q \Hubble}{a^{7/2}} \frac{d \delta \lambda}{d a} + \frac{2 \Phi Q m}{a^3} \right).
\label{aqw}
\end{align}

Thus, the perturbed energy density is identified as

\begin{equation}\label{pert}
\delta \mu = a^{-2} \delta T_{00}.
\end{equation}

These perturbation equations set the foundation for analyzing the growth of cosmic structures within the relativistic superfluid dark matter framework. In the subsequent section, we will demonstrate how these results reproduce those obtained in \cite{banerjee2020growth} before proceeding with a comprehensive relativistic treatment.
\section{Structure Formation}

The background Friedmann equation in cosmic time \(t\) is
\begin{equation}
H^2 = \frac{8\pi G}{3}\,\bar{\mu},
\end{equation}
where \(H = \frac{\dot{a}}{a}\) is the Hubble parameter and \(\bar{\mu}\) denotes the background energy density. Evaluating this equation today (\(a_0 \equiv 1\)) fixes the present-day background density,
\begin{equation}
\bar{\mu}_0 = \frac{3}{8\pi G} H_0^2.
\end{equation}

If the superfluid dark matter component dominates the background (\(\bar{\mu}_0 = m Q\)), one obtains
\begin{equation}\label{rdf}
m Q = \frac{3}{8\pi} H_0^2 M_{\mathrm{Pl}}^2.
\end{equation}
Using the reduced Planck mass \(\overline{M}_{\mathrm{Pl}}^2 = \frac{1}{8\pi G}\), this can equivalently be written as
\begin{equation}
m Q = 3 \, \overline{M}_{\mathrm{Pl}}^2 H_0^2.
\end{equation}
Numerically, taking \(H_0 \simeq 2.13 \times 10^{-33}\,\mathrm{eV}\) and \(M_{\mathrm{Pl}} \simeq 1.22 \times 10^{19}\,\mathrm{GeV}\) leads to
\begin{equation}
m Q \simeq 8.1 \times 10^{-47}\,\mathrm{GeV}^4.
\end{equation}

For a characteristic scale \(\Lambda_c = \Lambda = 1\,\mathrm{eV}\), the hierarchy
\begin{equation}
\frac{m^2 Q^2}{\Lambda_c^8} \sim 6.6 \times 10^{-21} \ll 1
\end{equation}
ensures that terms proportional to \(\Lambda\) in equations (\ref{ovlt}), (\ref{awr}), and (\ref{aqw}) can be safely neglected. Under this approximation, equation (\ref{ovlt}) reduces to
\begin{equation}\label{eq:delta_rho_small_lambda}
\frac{d^2 \delta \rho}{d a^2} + \frac{5}{2a} \frac{d \delta \rho}{d a} - \frac{2 m \sqrt{\frac{Q}{m}}}{H_0} \frac{d \delta \lambda}{d a} - \frac{a}{H_0^2} (2 m^2 - k^2) \delta \rho - \Phi \sqrt{\frac{Q}{m}} \left[ \frac{3}{a^{7/2}} - \frac{2}{a^{1/2}} \left( \frac{m}{H_0} \right)^2 \right] = 0,
\end{equation}
while (\ref{awr}) simplifies to
\begin{equation}\label{eq:delta_lambda_small_lambda}
\frac{d^2 \delta \lambda}{d a^2} - \frac{1}{2a} \frac{d \delta \lambda}{d a} - \frac{a}{H_0^2} (2 m^2 - k^2) \delta \lambda + \frac{2 m}{H_0 \sqrt{\frac{Q}{m}}} a^2 \frac{d \delta \rho}{d a} + \frac{3 m}{H_0 \sqrt{\frac{Q}{m}}} a \delta \rho + 6 \Phi \frac{m}{H_0} a^{-1/2} = 0.
\end{equation}

The perturbed mass density, given by (\ref{pert}), reduces to
\begin{equation}\label{eq:delta_mu_small_lambda}
\delta \mu = \frac{3 Q H_0}{a^{7/2}} \frac{d \delta \lambda}{d a} - \frac{9 H_0^2}{a^{7/2}} \frac{d \delta \rho}{d a} + \frac{2 m^2}{a^{3/2}} \sqrt{\frac{Q}{m}} \delta \rho + \frac{3 \Phi m Q}{a^3}.
\end{equation}

\subsection*{Low-\(k\) limit}
In the regime \(k \ll m\), where the gravitational potential \(\Phi\) evolves according to \eqref{poisson}, equation \eqref{eq:delta_rho_small_lambda} further simplifies to
\begin{equation}\label{eq:delta_rho_approx}
\frac{d^2 \delta \rho}{d a^2} + \frac{5}{2 a} \frac{d \delta \rho}{d a} - \frac{10}{3} \left( \frac{m}{H_0} \right)^2 a \delta \rho = 0,
\end{equation}
which admits the growing mode
\begin{equation}\label{aft}
\delta \rho(a) = \frac{A}{a^{3/2}} \exp\left[ a^{3/2} \frac{3}{2} \sqrt{\frac{10}{3}} \frac{m}{H_0} \right],
\end{equation}
with \(A\) an integration constant. Similarly, equation \eqref{eq:delta_lambda_small_lambda} reduces to
\begin{equation}\label{eq:delta_lambda_approx}
\frac{d^2 \delta \lambda}{d a^2} - \frac{13}{2 a} \frac{d \delta \lambda}{d a} - 2 \left( \frac{m}{H_0} \right)^2 a \delta \lambda = 0,
\end{equation}
with growing mode
\begin{equation}\label{afo}
\delta \lambda(a) = B \exp\left[ a^{3/2} \frac{3}{2} \sqrt{2} \frac{m}{H_0} \right] 
\left[ 8 \left( \frac{m}{H_0} \right)^3 a^3 - 18 \sqrt{2} \frac{m}{H_0} a^{3/2} + 27 \right],
\end{equation}
where \(B\) is an integration constant. These solutions grow extremely rapid as shown in Fig~\ref{fig:delta_oscillation}, and are therefore discarded as nonphysical.
\begin{figure}[h!]
    \centering
    \includegraphics[width=0.8\textwidth]{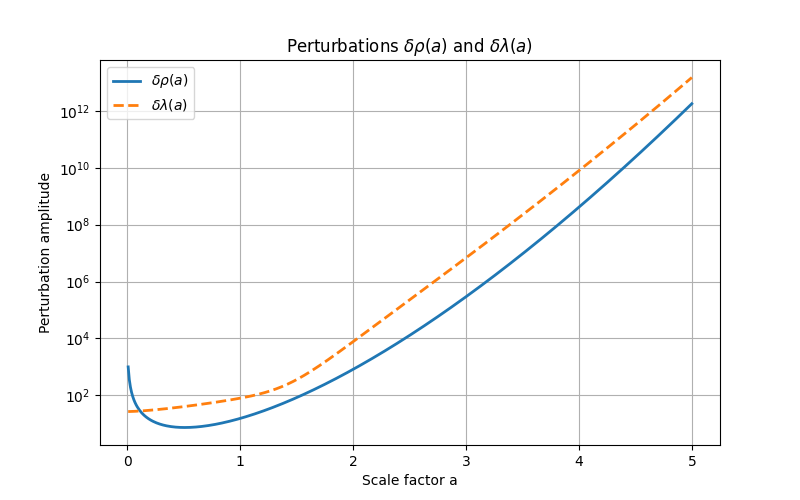}
     \captionsetup{justification=raggedright,singlelinecheck=false}
     \caption{Analytically approximated perturbations $\delta \lambda(a)$ and $\delta \rho(a)$ derived from \eqref{afo} and \eqref{aft} respectively. The figure illustrates how both perturbations grow rapidly with the expansion of the Universe, highlighting their relative behavior across the scale factor range.}
\label{fig:delta_oscillation}
\end{figure}
\subsection*{High-\(k\) limit}
In the complementary regime \(k \gg m\), equation \eqref{eq:delta_rho_small_lambda} becomes
\begin{equation}\label{eq:delta_rho_high_k}
\frac{d^2 \delta \rho}{d a^2} + \frac{5}{2 a} \frac{d \delta \rho}{d a} + \frac{k^2}{H_0^2} \delta \rho = 0,
\end{equation}
with oscillatory solution
\begin{equation}\label{laj}
\delta \rho(a) = A a^{-3/2} \cos \left( \frac{2 k}{3 H_0} a^{3/2} \right).
\end{equation}
Similarly, equation \eqref{eq:delta_lambda_small_lambda} reduces to
\begin{equation}\label{eq:delta_lambda_high_k}
\frac{d^2 \delta \lambda}{d a^2} - \frac{1}{2 a} \frac{d \delta \lambda}{d a} + \frac{k^2 a}{H_0^2} \delta \lambda = 0,
\end{equation}
which admits
\begin{equation}\label{lak}
\delta \lambda(a) = D \cos\left( \frac{2 k}{3 H_0} a^{3/2} - x \right),
\end{equation}
where \(D = \sqrt{B^2 + C^2}\) and \(x = \tan^{-1}(\frac{C}{B})\).
\begin{figure}[h!]
    \centering
    \includegraphics[width=0.8\textwidth]{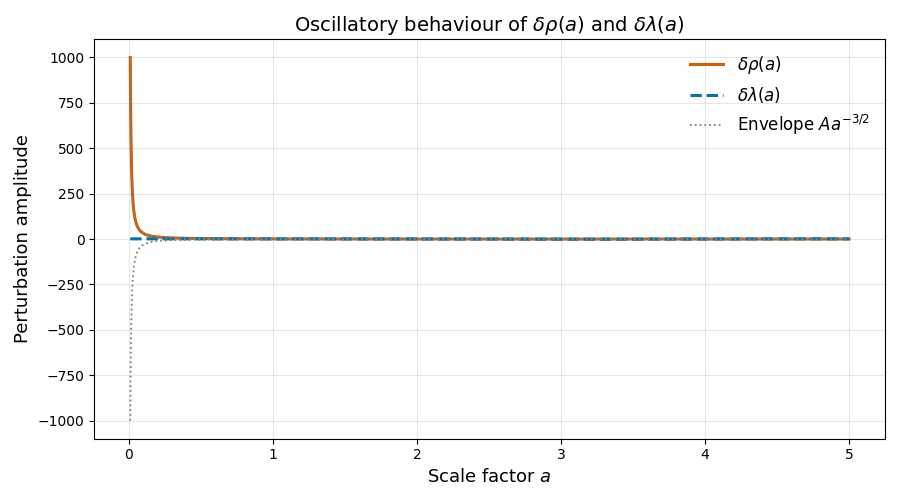}
     \captionsetup{justification=raggedright,singlelinecheck=false}
     \caption{Small scale factor ($a \ll 1$) behavior of the perturbations \eqref{laj} and \eqref{lak}. Both $\delta\rho(a)$ (orange curve) and $\delta\lambda(a)$ (blue curve) oscillate slowly with nearly constant amplitudes, showing that the information in the perturbations is fully visible in this regime before the onset of rapid oscillations and decay at large $a$.}
\label{fig:delta_oscillation1}
\end{figure}
\begin{figure}[h!]
    \centering
    \includegraphics[width=0.8\textwidth]{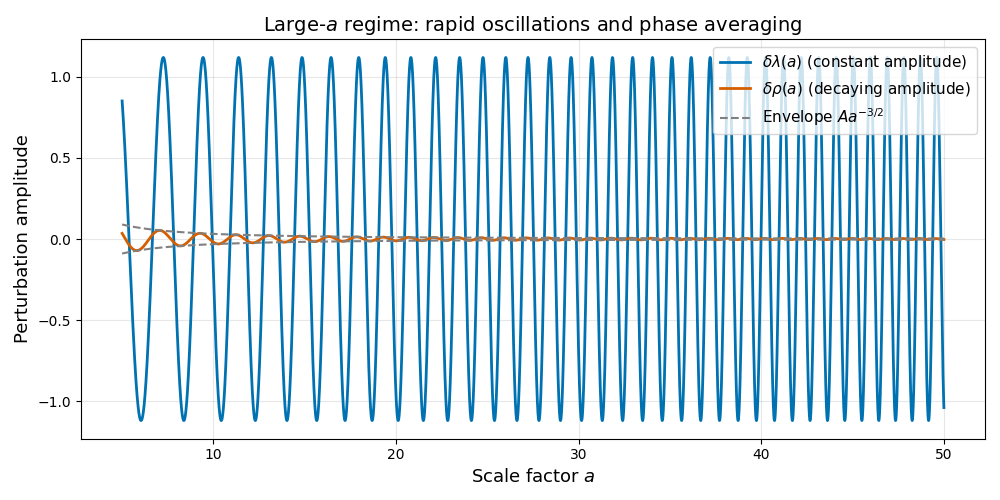}
     \captionsetup{justification=raggedright,singlelinecheck=false}
     \caption{Large scale factor ($a \gg 1$) behavior of the perturbations \eqref{laj} and \eqref{lak}. The orange curve shows $\delta\rho(a)$, whose amplitude decays as $A a^{-3/2}$ (gray dashed envelope) while oscillating rapidly. The blue curve shows $\delta\lambda(a)$, which oscillates at constant amplitude $D$ with a phase shift $x = \arctan(C/B)$. At large $a$, the rapid oscillations of $\delta\rho(a)$ effectively average out, illustrating how information can become hidden in high-frequency, decaying perturbations.}
\label{fig:delta_oscillation2}
\end{figure}
The behavior of \eqref{laj} and \eqref{lak} is displayed in Figs.~\ref{fig:delta_oscillation1}---\ref{fig:delta_oscillation2}. Substituting these solutions into \eqref{eq:delta_mu_small_lambda} gives the total density perturbation
\begin{multline}\label{delmu}
\delta \mu(a) = \frac{27}{4} H_0^2 \sqrt{\frac{Q}{m}} A a^{-\frac{11}{2}} \cos\left( \frac{2 k}{3 H_0} a^{\frac{3}{2}} \right) 
+ \frac{9}{2} \sqrt{\frac{Q}{m}} H_0 k A a^{-\frac{9}{2}} \sin\left( \frac{2 k}{3 H_0} a^{\frac{3}{2}} \right) 
\\- 3 Q k D a^{-3} \sin\left( \frac{2 k}{3 H_0} a^{\frac{3}{2}} - x \right) + 2 m^2 \sqrt{\frac{Q}{m}} A \cos\left( \frac{2 k}{3 H_0} a^{\frac{3}{2}} \right).
\end{multline}
Defining the density contrast \(\delta = \frac{\delta \mu}{\bar{\mu}}\) with \(\bar{\mu} = m Q a^{-3}\) leads to
\begin{multline}\label{eq:density_contrast}
\delta(a) = \frac{27}{4} \frac{H_0^2}{m Q} \sqrt{\frac{Q}{m}} A a^{-\frac{5}{2}} \cos\left( \frac{2 k}{3 H_0} a^{\frac{3}{2}} \right) 
+ \frac{9}{2} \sqrt{\frac{Q}{m}} \frac{H_0}{m Q} k A a^{-\frac{3}{2}} \sin\left( \frac{2 k}{3 H_0} a^{\frac{3}{2}} \right) 
\\- 3 \frac{k}{m} D \sin\left( \frac{2 k}{3 H_0} a^{\frac{3}{2}} - x \right) 
+ 2 a^{-3} \sqrt{\frac{m}{Q}} A \cos\left( \frac{2 k}{3 H_0} a^{\frac{3}{2}} \right).
\end{multline}
At late times, all terms decay except the third, giving the approximation
\begin{equation}\label{eq:density_contrast_approx}
\delta(a) \approx \frac{3 k}{m} D \sin\left( -\frac{2 k}{3 H_0} a^{3/2} + x \right).
\end{equation}
\begin{figure}[h!]
    \centering
\includegraphics[width=0.8\textwidth]{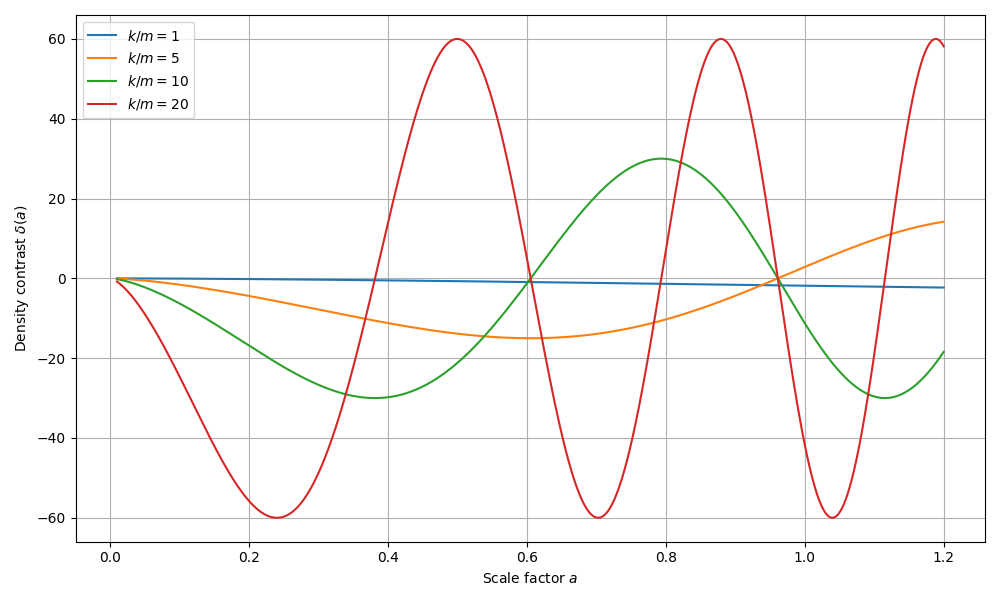}
\captionsetup{justification=raggedright,singlelinecheck=false}    
\caption{Oscillatory behavior of the density contrast $\delta(a)$ in the regime $k \gg m$, as described by Equation \eqref{eq:density_contrast_approx}. The amplitude remains constant while the frequency increases with the scale factor, indicating suppressed structure growth in the absence of baryonic coupling.}
\label{fig:delta_oscillation}
\end{figure}
Averaging over a time interval \(\Delta T\) yields a constant mean,
\begin{equation}
\langle \delta(t) \rangle = \frac{3kD}{m \Delta T} \int_0^{\Delta T} \sin\left( -\frac{2k}{3 H_0} a(t)^{3/2} + x \right) \, dt = \text{constant},
\end{equation}
in agreement with \cite{banerjee2020growth} when baryonic couplings are neglected, though differing from the growth predicted in \(\Lambda\)CDM \cite{bertschinger1998simulations, brandenberger1995formation, peacock2002introduction}. This confirms that, in the weak-field limit, structure formation is suppressed for superfluid dark matter in the absence of baryonic interactions. Next, we consider a relativistic extension of this first-order linear theory.
\section{Emergence of \(\Lambda\)CDM Behavior in the Large Scalar Mass Regime}

To investigate the behavior of structure formation when the scalar field mass is large, we consider power-law solutions of the form
\begin{equation}\label{tarf}
\delta\rho(a) = \delta\rho_0\, a^{m_\rho}, \qquad 
\delta\lambda(a) = \delta\lambda_0\, a^{m_\lambda}, \qquad 
\Phi(a) = \Phi_0,
\end{equation}
where \(\Phi\) is taken to be approximately constant on sub-Hubble scales (\(k \gg H\)) during the matter-dominated era, a valid assumption in this regime. Dividing equation~\eqref{eq:delta_mu_small_lambda} by the background energy density \(\bar{\mu} = m Q a^{-3}\), the density contrast can be expressed as
\begin{equation}
\delta(a) = \frac{\delta \mu}{\bar{\mu}} = \frac{3 H_0 m_\lambda \delta \lambda_0}{m} \, a^{m_\lambda - \frac{3}{2}} 
- \frac{9 H_0^{2} m_\rho \delta \rho_0}{m Q} \, a^{m_\rho - \frac{3}{2}} 
+ 2 m \frac{\sqrt{m}}{\sqrt{Q}} \, \delta \rho_0 \, a^{m_\rho + \frac{3}{2}} 
+ 3 \Phi_0.
\end{equation}
In the large mass limit corresponding to equation~\eqref{afg}, the density contrast simplifies to\footnote{See the Appendix for the full derivation.}
\begin{equation}\label{sawr}
\delta(a) \propto 2 \sqrt{m Q}\, \delta \rho_0 + 3 \Phi_0= \text{constant}.
\end{equation}
Hence, structure growth is suppressed, paralleling the behavior found in the weak-field limit. The evolution of baryonic density perturbations in the presence of a dark sector component follows the standard perturbation equations in the conformal Newtonian gauge. Assuming that baryons and the complex scalar field are only coupled gravitationally, with no direct interactions or energy exchange between them, the total density contrast is given by the weighted sum
\begin{equation}\label{axd}
\delta_{\text{tot}} = \frac{\bar{\mu}_b}{\bar{\mu}_b + \bar{\mu}} \, \delta_b + \frac{\bar{\mu}}{\bar{\mu}_b + \bar{\mu}} \, \delta,
\end{equation}
where \(\delta_b\) is the baryon density contrast and \(\bar{\mu}_b\) is the energy density of baryons~\cite{Ma:1995ey, Dodelson:2003}.

\subsection{Total Density Contrast and CDM Limit}
Inserting \(\bar{\mu} = m Q a^{-3}\) and \(\bar{\mu}_b=\mu_{b0}a^{-3}\) into equation~\eqref{axd} lead us to
\begin{equation}
\delta_{\text{tot}} = \frac{\mu_{b0} a^{-3}}{\mu_{b0} a^{-3} + m Q a^{-3}} \delta_b + \frac{m Q a^{-3}}{\mu_{b0} a^{-3} + m Q a^{-3}} \delta.
\end{equation}
Since \(a^{-3}\) cancels out we get,
\begin{equation}
\delta_{\text{tot}} = \frac{\mu_{b0}}{\mu_{b0} + m Q} \delta_b + \frac{m Q}{\mu_{b0} + m Q} \delta.
\end{equation}

Solving the linearized perturbation equations during matter domination reveals a growing mode for baryons scaling as \(\delta_b(a) \propto a\)~\cite{Ma_Bertschinger_1995, Dodelson_2003}. In the large mass limit, the complex scalar field energy density dominates the total matter content at late times due to its \(a^{-3}\) scaling. Meanwhile, the scalar field density contrast \(\delta\) approaches a constant value, as shown in equation~\eqref{sawr}. Consequently, the total density contrast reduces effectively to the baryon density contrast weighted by their relative contributions. Since \(\delta_b\) grows linearly with scale factor during matter domination, the total density contrast \(\delta_{\text{tot}}\) inherits this linear growth, thereby reproducing the canonical cold dark matter behavior characteristic of \(\Lambda\)CDM cosmology.

\section{Conclusion}
In this work, we have explored the relativistic formulation of the dark matter superfluidity paradigm and examined its implications for cosmological structure formation. Starting from the effective field theory Lagrangian, we constructed the background dynamics in a Friedmann–Lemaître–Robertson–Walker spacetime and performed a linear perturbation analysis to trace the evolution of scalar perturbations in both the phonon and density sectors. Our findings show that in the weak-field regime and in the absence of baryonic couplings, the growth of density perturbations is strongly suppressed, consistent with earlier results in the literature. Specifically, in the limit \(k \gg m\), the density contrast exhibits oscillatory behavior with a constant amplitude envelope, indicating that large-scale structures do not grow during this phase. This suppression arises due to the ultra-light mass of the scalar field and the absence of coupling to baryons, which otherwise mediate the MOND-like force on galactic scales. 
\\
\\
Furthermore, we demonstrated that in the limit of a large scalar field mass \(m \gg H_0\), the theory reduces to standard $\Lambda$CDM behavior. In this regime, the scalar degree of freedom effectively decouples, and the dominant growing mode of the density contrast scales linearly with the scale factor, \(\delta \propto a\), as expected in the matter-dominated era of conventional cosmology. This key result confirms that the relativistic dark matter superfluid model recovers the successful predictions of $\Lambda$CDM on large scales, while simultaneously accommodating MOND-like dynamics at galactic scales. Overall, our analysis reinforces the viability of dark matter superfluidity as a unified framework capable of interpolating between modified gravity effects in galaxies and standard cosmological behavior in the early Universe. Future work will aim to include baryon couplings explicitly in the relativistic perturbation equations and to investigate potential observational signatures that may distinguish this scenario from other dark sector models.
\section*{Acknowledgments}
I would like to sincerely thank Dr.~Julien Larena for suggesting this problem as a topic for my master's thesis and for his valuable guidance throughout its development.
\appendix
\section{Power-Law Solutions for Constant \(\Phi\)}

In this appendix, we present a controlled analysis of the coupled linear perturbation equations for the superfluid dark matter system, under the assumption of a constant gravitational potential \(\Phi\). We seek power-law solutions for the perturbations without invoking any late-time approximation, and work within a controlled large-mass approximation. The analysis is performed in the limit of small \(\Lambda\), allowing the linearized equations to be simplified.

\subsection{Linearized Perturbation Equations}

In the small-\(\Lambda\) limit, the linearized perturbation equations take the form
\begin{align}
\frac{d^2 \delta \rho}{d a^2} + \frac{5}{2a} \frac{d \delta \rho}{d a} 
- \frac{2 m \sqrt{\frac{Q}{m}}}{H_0} \frac{d \delta \lambda}{d a} 
- \frac{a}{H_0^2} (2 m^2 - k^2) \delta \rho 
- \Phi \sqrt{\frac{Q}{m}} \left[ \frac{3}{a^{7/2}} - 2 \frac{m^2}{H_0^2} a^{-1/2} \right] &= 0, \label{eq:rho_appendix} \\
\frac{d^2 \delta \lambda}{d a^2} - \frac{1}{2a} \frac{d \delta \lambda}{d a} 
- \frac{a}{H_0^2} (2 m^2 - k^2) \delta \lambda 
+ \frac{2 m}{H_0 \sqrt{\frac{Q}{m}}} a^2 \frac{d \delta \rho}{d a} 
+ \frac{3 m}{H_0 \sqrt{\frac{Q}{m}}} a \delta \rho 
+ 6 \Phi \frac{m}{H_0} a^{-1/2} &= 0. \label{eq:lambda_appendix}
\end{align}

\subsection{Power-Law Ansatz}

We assume a power-law form for the perturbations,
\begin{equation}\label{eq:ansatz_appendix}
\delta \rho(a) = \delta \rho_0 a^{m_\rho}, \qquad
\delta \lambda(a) = \delta \lambda_0 a^{m_\lambda},
\end{equation}
with constant amplitudes \(\delta \rho_0, \delta \lambda_0\) and exponents \(m_\rho, m_\lambda\).  
The derivatives are
\begin{align*}
\frac{d \delta \rho}{d a} &= m_\rho \delta \rho_0 a^{m_\rho - 1}, & 
\frac{d^2 \delta \rho}{d a^2} &= m_\rho (m_\rho - 1) \delta \rho_0 a^{m_\rho - 2}, \\
\frac{d \delta \lambda}{d a} &= m_\lambda \delta \lambda_0 a^{m_\lambda - 1}, & 
\frac{d^2 \delta \lambda}{d a^2} &= m_\lambda (m_\lambda - 1) \delta \lambda_0 a^{m_\lambda - 2}.
\end{align*}

\subsection{Substitution into the Equations}

Substituting the ansatz into Eqs.~\eqref{eq:rho_appendix}--\eqref{eq:lambda_appendix} gives
\begin{align}
& \left(m_\rho^2 + \frac{3}{2} m_\rho \right) \delta \rho_0 a^{m_\rho - 2} 
- \frac{2 m \sqrt{\frac{Q}{m}}}{H_0} m_\lambda \delta \lambda_0 a^{m_\lambda - 1} 
- \frac{2 m^2 - k^2}{H_0^2} \delta \rho_0 a^{m_\rho + 1} \nonumber \\
& \quad - \Phi \sqrt{\frac{Q}{m}} \left[ 3 a^{-7/2} - 2 \frac{m^2}{H_0^2} a^{-1/2} \right] = 0, \\
& \left(m_\lambda^2 - \frac{3}{2} m_\lambda \right) \delta \lambda_0 a^{m_\lambda - 2} 
- \frac{2 m^2 - k^2}{H_0^2} \delta \lambda_0 a^{m_\lambda + 1} 
+ \frac{2 m m_\rho + 3 m}{H_0 \sqrt{Q/m}} \delta \rho_0 a^{m_\rho + 1} 
+ 6 \Phi \frac{m}{H_0} a^{-1/2} = 0.
\end{align}

\subsection{Homogeneous Solution}

An exact power-law solution valid for all \(a\) does not exist, because the terms in the homogeneous equations carry different powers of \(a\). Nevertheless, in the large-mass limit \(m \gg H_0, k\), the terms proportional to \(\frac{2 m^2}{H_0^2}\) dominate, while derivative terms are suppressed by factors of \(\frac{H_0}{m} \ll 1\). Performing a leading-order approximation, we find an approximately constant homogeneous mode, with amplitudes related by
\begin{equation}
\delta \rho_0 \simeq - \frac{H_0 \sqrt{Q/m}}{m} m_h \, \delta \lambda_0,
\end{equation}
where \(m_h \ll 1\) is the effective small exponent of the nearly constant mode. The homogeneous solution is then
\begin{equation}
\delta \rho(a) \simeq \delta \rho_0, \quad \delta \lambda(a) \simeq \delta \lambda_0.
\end{equation}

\subsection{Forced Solution}

The \(\Phi\)-dependent terms act as a source, producing approximately constant contributions at leading order:
\begin{equation}
\delta \rho_\Phi \simeq 2 \Phi \sqrt{m Q}, \qquad \delta \lambda_\Phi \simeq 3 \Phi.
\end{equation}

The full solution is therefore
\begin{equation}
\delta \rho(a) = \delta \rho_0 + \delta \rho_\Phi, \quad
\delta \lambda(a) = \delta \lambda_0 + \delta \lambda_\Phi.
\end{equation}

\subsection{Total Density Contrast}

The total density perturbation is
\begin{equation}
\delta \mu(a) = \frac{3 Q H_0}{a^{7/2}} \frac{d \delta \lambda}{d a} - \frac{9 H_0^2}{a^{7/2}} \frac{d \delta \rho}{d a} + \frac{2 m^2}{a^{3/2}} \sqrt{\frac{Q}{m}}\, \delta \rho + \frac{3 \Phi m Q}{a^3}.
\end{equation}

Substituting the approximate homogeneous and forced solutions, and dividing by the background density \(\bar{\mu} = m Q a^{-3}\), we obtain
\begin{equation}
\delta(a) = \frac{\delta \mu(a)}{\bar{\mu}(a)} \simeq 2 \sqrt{m Q}\, \delta \rho_0 + 3 \Phi,
\end{equation}
which is independent of the scale factor at leading order in the large-mass limit.

\noindent
Hence, the density contrast is controlled by the initial amplitudes and the gravitational potential. Residual variations with \(a\) arising from derivative terms are subleading in \(\frac{H_0}{m}\) and can be safely neglected at leading order.
\bibliographystyle{apsrev4-2}
\bibliography{main}

\begin{thebibliography}{31}%
\makeatletter
\providecommand \@ifxundefined [1]{%
 \@ifx{#1\undefined}
}%
\providecommand \@ifnum [1]{%
 \ifnum #1\expandafter \@firstoftwo
 \else \expandafter \@secondoftwo
 \fi
}%
\providecommand \@ifx [1]{%
 \ifx #1\expandafter \@firstoftwo
 \else \expandafter \@secondoftwo
 \fi
}%
\providecommand \natexlab [1]{#1}%
\providecommand \enquote  [1]{``#1''}%
\providecommand \bibnamefont  [1]{#1}%
\providecommand \bibfnamefont [1]{#1}%
\providecommand \citenamefont [1]{#1}%
\providecommand \href@noop [0]{\@secondoftwo}%
\providecommand \href [0]{\begingroup \@sanitize@url \@href}%
\providecommand \@href[1]{\@@startlink{#1}\@@href}%
\providecommand \@@href[1]{\endgroup#1\@@endlink}%
\providecommand \@sanitize@url [0]{\catcode `\\12\catcode `\$12\catcode `\&12\catcode `\#12\catcode `\^12\catcode `\_12\catcode `\%12\relax}%
\providecommand \@@startlink[1]{}%
\providecommand \@@endlink[0]{}%
\providecommand \url  [0]{\begingroup\@sanitize@url \@url }%
\providecommand \@url [1]{\endgroup\@href {#1}{\urlprefix }}%
\providecommand \urlprefix  [0]{URL }%
\providecommand \Eprint [0]{\href }%
\providecommand \doibase [0]{https://doi.org/}%
\providecommand \selectlanguage [0]{\@gobble}%
\providecommand \bibinfo  [0]{\@secondoftwo}%
\providecommand \bibfield  [0]{\@secondoftwo}%
\providecommand \translation [1]{[#1]}%
\providecommand \BibitemOpen [0]{}%
\providecommand \bibitemStop [0]{}%
\providecommand \bibitemNoStop [0]{.\EOS\space}%
\providecommand \EOS [0]{\spacefactor3000\relax}%
\providecommand \BibitemShut  [1]{\csname bibitem#1\endcsname}%
\let\auto@bib@innerbib\@empty
\bibitem [{\citenamefont {Peebles}(2015)}]{peebles2015dark}%
  \BibitemOpen
  \bibfield  {author} {\bibinfo {author} {\bibfnamefont {P.~J.~E.}\ \bibnamefont {Peebles}},\ }\href@noop {} {\bibfield  {journal} {\bibinfo  {journal} {Proceedings of the National Academy of Sciences}\ }\textbf {\bibinfo {volume} {112}},\ \bibinfo {pages} {12246} (\bibinfo {year} {2015})}\BibitemShut {NoStop}%
\bibitem [{\citenamefont {Ellis}\ \emph {et~al.}(2012)\citenamefont {Ellis}, \citenamefont {Maartens},\ and\ \citenamefont {MacCallum}}]{ellis2012relativistic}%
  \BibitemOpen
  \bibfield  {author} {\bibinfo {author} {\bibfnamefont {G.~F.}\ \bibnamefont {Ellis}}, \bibinfo {author} {\bibfnamefont {R.}~\bibnamefont {Maartens}},\ and\ \bibinfo {author} {\bibfnamefont {M.~A.}\ \bibnamefont {MacCallum}},\ }\href@noop {} {\emph {\bibinfo {title} {Relativistic cosmology}}}\ (\bibinfo  {publisher} {Cambridge University Press},\ \bibinfo {year} {2012})\BibitemShut {NoStop}%
\bibitem [{\citenamefont {Rubin}\ and\ \citenamefont {Ford~Jr}(1970)}]{rubin1970rotation}%
  \BibitemOpen
  \bibfield  {author} {\bibinfo {author} {\bibfnamefont {V.~C.}\ \bibnamefont {Rubin}}\ and\ \bibinfo {author} {\bibfnamefont {W.~K.}\ \bibnamefont {Ford~Jr}},\ }\href@noop {} {\bibfield  {journal} {\bibinfo  {journal} {Astrophysical Journal, vol. 159, p. 379}\ }\textbf {\bibinfo {volume} {159}},\ \bibinfo {pages} {379} (\bibinfo {year} {1970})}\BibitemShut {NoStop}%
\bibitem [{\citenamefont {Freeman}(2001)}]{freeman2001dark}%
  \BibitemOpen
  \bibfield  {author} {\bibinfo {author} {\bibfnamefont {K.}~\bibnamefont {Freeman}},\ }in\ \href@noop {} {\emph {\bibinfo {booktitle} {Encyclopedia of Astronomy \& Astrophysics}}}\ (\bibinfo  {publisher} {CRC Press},\ \bibinfo {year} {2001})\ pp.\ \bibinfo {pages} {1--3}\BibitemShut {NoStop}%
\bibitem [{\citenamefont {McGaugh}\ \emph {et~al.}(2000)\citenamefont {McGaugh}, \citenamefont {Schombert}, \citenamefont {Bothun},\ and\ \citenamefont {De~Blok}}]{mcgaugh2000baryonic}%
  \BibitemOpen
  \bibfield  {author} {\bibinfo {author} {\bibfnamefont {S.~S.}\ \bibnamefont {McGaugh}}, \bibinfo {author} {\bibfnamefont {J.~M.}\ \bibnamefont {Schombert}}, \bibinfo {author} {\bibfnamefont {G.~D.}\ \bibnamefont {Bothun}},\ and\ \bibinfo {author} {\bibfnamefont {W.}~\bibnamefont {De~Blok}},\ }\href@noop {} {\bibfield  {journal} {\bibinfo  {journal} {The Astrophysical Journal}\ }\textbf {\bibinfo {volume} {533}},\ \bibinfo {pages} {L99} (\bibinfo {year} {2000})}\BibitemShut {NoStop}%
\bibitem [{\citenamefont {McGaugh}\ \emph {et~al.}(2016)\citenamefont {McGaugh}, \citenamefont {Lelli},\ and\ \citenamefont {Schombert}}]{mcgaugh2016radial}%
  \BibitemOpen
  \bibfield  {author} {\bibinfo {author} {\bibfnamefont {S.~S.}\ \bibnamefont {McGaugh}}, \bibinfo {author} {\bibfnamefont {F.}~\bibnamefont {Lelli}},\ and\ \bibinfo {author} {\bibfnamefont {J.~M.}\ \bibnamefont {Schombert}},\ }\href@noop {} {\bibfield  {journal} {\bibinfo  {journal} {Physical Review Letters}\ }\textbf {\bibinfo {volume} {117}},\ \bibinfo {pages} {201101} (\bibinfo {year} {2016})}\BibitemShut {NoStop}%
\bibitem [{\citenamefont {Governato}\ \emph {et~al.}(2010)\citenamefont {Governato} \emph {et~al.}}]{governato2010bulgeless}%
  \BibitemOpen
  \bibfield  {author} {\bibinfo {author} {\bibfnamefont {F.}~\bibnamefont {Governato}} \emph {et~al.},\ }\href@noop {} {\bibfield  {journal} {\bibinfo  {journal} {Nature}\ }\textbf {\bibinfo {volume} {463}},\ \bibinfo {pages} {203} (\bibinfo {year} {2010})}\BibitemShut {NoStop}%
\bibitem [{\citenamefont {Di~Cintio}\ \emph {et~al.}(2014)\citenamefont {Di~Cintio} \emph {et~al.}}]{dicintio2014core}%
  \BibitemOpen
  \bibfield  {author} {\bibinfo {author} {\bibfnamefont {A.}~\bibnamefont {Di~Cintio}} \emph {et~al.},\ }\href@noop {} {\bibfield  {journal} {\bibinfo  {journal} {Monthly Notices of the Royal Astronomical Society}\ }\textbf {\bibinfo {volume} {437}},\ \bibinfo {pages} {415} (\bibinfo {year} {2014})}\BibitemShut {NoStop}%
\bibitem [{\citenamefont {Del~Popolo}\ and\ \citenamefont {Le~Delliou}(2017)}]{trujillo2016cold}%
  \BibitemOpen
  \bibfield  {author} {\bibinfo {author} {\bibfnamefont {A.}~\bibnamefont {Del~Popolo}}\ and\ \bibinfo {author} {\bibfnamefont {M.}~\bibnamefont {Le~Delliou}},\ }\href@noop {} {\bibfield  {journal} {\bibinfo  {journal} {Galaxies}\ }\textbf {\bibinfo {volume} {5}},\ \bibinfo {pages} {17} (\bibinfo {year} {2017})}\BibitemShut {NoStop}%
\bibitem [{\citenamefont {Famaey}\ and\ \citenamefont {McGaugh}(2012)}]{famaey2012modified}%
  \BibitemOpen
  \bibfield  {author} {\bibinfo {author} {\bibfnamefont {B.}~\bibnamefont {Famaey}}\ and\ \bibinfo {author} {\bibfnamefont {S.~S.}\ \bibnamefont {McGaugh}},\ }\href@noop {} {\bibfield  {journal} {\bibinfo  {journal} {Living reviews in relativity}\ }\textbf {\bibinfo {volume} {15}},\ \bibinfo {pages} {1} (\bibinfo {year} {2012})}\BibitemShut {NoStop}%
\bibitem [{\citenamefont {Milgrom}(2020)}]{milgrom2020mondreview}%
  \BibitemOpen
  \bibfield  {author} {\bibinfo {author} {\bibfnamefont {M.}~\bibnamefont {Milgrom}},\ }\href@noop {} {\bibfield  {journal} {\bibinfo  {journal} {Canadian Journal of Physics}\ }\textbf {\bibinfo {volume} {98}},\ \bibinfo {pages} {379} (\bibinfo {year} {2020})}\BibitemShut {NoStop}%
\bibitem [{\citenamefont {Milgrom}(2008)}]{milgrom2008mond}%
  \BibitemOpen
  \bibfield  {author} {\bibinfo {author} {\bibfnamefont {M.}~\bibnamefont {Milgrom}},\ }\href@noop {} {\bibfield  {journal} {\bibinfo  {journal} {arXiv preprint arXiv:0801.3133}\ } (\bibinfo {year} {2008})}\BibitemShut {NoStop}%
\bibitem [{\citenamefont {Bekenstein}\ and\ \citenamefont {Milgrom}(1984)}]{bekenstein1984does}%
  \BibitemOpen
  \bibfield  {author} {\bibinfo {author} {\bibfnamefont {J.}~\bibnamefont {Bekenstein}}\ and\ \bibinfo {author} {\bibfnamefont {M.}~\bibnamefont {Milgrom}},\ }\href@noop {} {\bibfield  {journal} {\bibinfo  {journal} {Astrophysical Journal, Part 1 (ISSN 0004-637X), vol. 286, Nov. 1, 1984, p. 7-14. Research supported by the MINERVA Foundation.}\ }\textbf {\bibinfo {volume} {286}},\ \bibinfo {pages} {7} (\bibinfo {year} {1984})}\BibitemShut {NoStop}%
\bibitem [{\citenamefont {Berezhiani}\ and\ \citenamefont {Khoury}(2015{\natexlab{a}})}]{r1}%
  \BibitemOpen
  \bibfield  {author} {\bibinfo {author} {\bibfnamefont {L.}~\bibnamefont {Berezhiani}}\ and\ \bibinfo {author} {\bibfnamefont {J.}~\bibnamefont {Khoury}},\ }\href@noop {} {\bibfield  {journal} {\bibinfo  {journal} {Physical Review D}\ }\textbf {\bibinfo {volume} {92}},\ \bibinfo {pages} {103510} (\bibinfo {year} {2015}{\natexlab{a}})}\BibitemShut {NoStop}%
\bibitem [{\citenamefont {Berezhiani}\ \emph {et~al.}(2018)\citenamefont {Berezhiani}, \citenamefont {Famaey},\ and\ \citenamefont {Khoury}}]{berezhiani2018phenomenological}%
  \BibitemOpen
  \bibfield  {author} {\bibinfo {author} {\bibfnamefont {L.}~\bibnamefont {Berezhiani}}, \bibinfo {author} {\bibfnamefont {B.}~\bibnamefont {Famaey}},\ and\ \bibinfo {author} {\bibfnamefont {J.}~\bibnamefont {Khoury}},\ }\href@noop {} {\bibfield  {journal} {\bibinfo  {journal} {Journal of Cosmology and Astroparticle Physics}\ }\textbf {\bibinfo {volume} {2018}}\bibinfo  {number} { (09)},\ \bibinfo {pages} {021}}\BibitemShut {NoStop}%
\bibitem [{\citenamefont {Khoury}(2015)}]{khoury2015dark}%
  \BibitemOpen
\bibfield  {number} {  }\bibfield  {author} {\bibinfo {author} {\bibfnamefont {J.}~\bibnamefont {Khoury}},\ }\href@noop {} {\bibfield  {journal} {\bibinfo  {journal} {arXiv preprint arXiv:1506.06640}\ } (\bibinfo {year} {2015})}\BibitemShut {NoStop}%
\bibitem [{\citenamefont {Berezhiani}\ and\ \citenamefont {Khoury}(2015{\natexlab{b}})}]{berezhiani2015theory}%
  \BibitemOpen
  \bibfield  {author} {\bibinfo {author} {\bibfnamefont {L.}~\bibnamefont {Berezhiani}}\ and\ \bibinfo {author} {\bibfnamefont {J.}~\bibnamefont {Khoury}},\ }\href@noop {} {\bibfield  {journal} {\bibinfo  {journal} {Physical Review D}\ }\textbf {\bibinfo {volume} {92}},\ \bibinfo {pages} {103510} (\bibinfo {year} {2015}{\natexlab{b}})}\BibitemShut {NoStop}%
\bibitem [{\citenamefont {Berezhiani}\ and\ \citenamefont {Khoury}(2015{\natexlab{c}})}]{berezhiani2015phenomenology}%
  \BibitemOpen
  \bibfield  {author} {\bibinfo {author} {\bibfnamefont {L.}~\bibnamefont {Berezhiani}}\ and\ \bibinfo {author} {\bibfnamefont {J.}~\bibnamefont {Khoury}},\ }\href@noop {} {\bibfield  {journal} {\bibinfo  {journal} {Physical Review D}\ }\textbf {\bibinfo {volume} {92}},\ \bibinfo {pages} {103509} (\bibinfo {year} {2015}{\natexlab{c}})}\BibitemShut {NoStop}%
\bibitem [{\citenamefont {Mistele}\ \emph {et~al.}(2022)\citenamefont {Mistele}, \citenamefont {McGaugh},\ and\ \citenamefont {Hossenfelder}}]{mistele2022galactic}%
  \BibitemOpen
  \bibfield  {author} {\bibinfo {author} {\bibfnamefont {T.}~\bibnamefont {Mistele}}, \bibinfo {author} {\bibfnamefont {S.}~\bibnamefont {McGaugh}},\ and\ \bibinfo {author} {\bibfnamefont {S.}~\bibnamefont {Hossenfelder}},\ }\href@noop {} {\bibfield  {journal} {\bibinfo  {journal} {Astronomy \& Astrophysics}\ }\textbf {\bibinfo {volume} {664}},\ \bibinfo {pages} {A40} (\bibinfo {year} {2022})}\BibitemShut {NoStop}%
\bibitem [{\citenamefont {Khoury}(2022)}]{khoury2022dark}%
  \BibitemOpen
  \bibfield  {author} {\bibinfo {author} {\bibfnamefont {J.}~\bibnamefont {Khoury}},\ }\href@noop {} {\bibfield  {journal} {\bibinfo  {journal} {SciPost Physics Lecture Notes}\ ,\ \bibinfo {pages} {042}} (\bibinfo {year} {2022})}\BibitemShut {NoStop}%
\bibitem [{\citenamefont {Son}(2002)}]{son2002low}%
  \BibitemOpen
  \bibfield  {author} {\bibinfo {author} {\bibfnamefont {D.}~\bibnamefont {Son}},\ }\href@noop {} {\bibfield  {journal} {\bibinfo  {journal} {arXiv preprint hep-ph/0204199}\ } (\bibinfo {year} {2002})}\BibitemShut {NoStop}%
\bibitem [{\citenamefont {Greiter}\ \emph {et~al.}(1989)\citenamefont {Greiter}, \citenamefont {Wilczek},\ and\ \citenamefont {Witten}}]{greiter1989hydrodynamic}%
  \BibitemOpen
  \bibfield  {author} {\bibinfo {author} {\bibfnamefont {M.}~\bibnamefont {Greiter}}, \bibinfo {author} {\bibfnamefont {F.}~\bibnamefont {Wilczek}},\ and\ \bibinfo {author} {\bibfnamefont {E.}~\bibnamefont {Witten}},\ }\href@noop {} {\bibfield  {journal} {\bibinfo  {journal} {Modern Physics Letters B}\ }\textbf {\bibinfo {volume} {3}},\ \bibinfo {pages} {903} (\bibinfo {year} {1989})}\BibitemShut {NoStop}%
\bibitem [{\citenamefont {Milgrom}(1983)}]{milgrom1983modification}%
  \BibitemOpen
  \bibfield  {author} {\bibinfo {author} {\bibfnamefont {M.}~\bibnamefont {Milgrom}},\ }\href@noop {} {\bibfield  {journal} {\bibinfo  {journal} {Astrophysical Journal, Part 1 (ISSN 0004-637X), vol. 270, July 15, 1983, p. 365-370. Research supported by the US-Israel Binational Science Foundation.}\ }\textbf {\bibinfo {volume} {270}},\ \bibinfo {pages} {365} (\bibinfo {year} {1983})}\BibitemShut {NoStop}%
\bibitem [{\citenamefont {Banerjee}\ \emph {et~al.}(2020)\citenamefont {Banerjee}, \citenamefont {Bera},\ and\ \citenamefont {Mota}}]{banerjee2020growth}%
  \BibitemOpen
  \bibfield  {author} {\bibinfo {author} {\bibfnamefont {S.}~\bibnamefont {Banerjee}}, \bibinfo {author} {\bibfnamefont {S.}~\bibnamefont {Bera}},\ and\ \bibinfo {author} {\bibfnamefont {D.~F.}\ \bibnamefont {Mota}},\ }\href@noop {} {\bibfield  {journal} {\bibinfo  {journal} {Journal of Cosmology and Astroparticle Physics}\ }\textbf {\bibinfo {volume} {2020}}\bibinfo  {number} { (07)},\ \bibinfo {pages} {034}}\BibitemShut {NoStop}%
\bibitem [{\citenamefont {Bertschinger}(1998)}]{bertschinger1998simulations}%
  \BibitemOpen
\bibfield  {number} {  }\bibfield  {author} {\bibinfo {author} {\bibfnamefont {E.}~\bibnamefont {Bertschinger}},\ }\href@noop {} {\bibfield  {journal} {\bibinfo  {journal} {Annual Review of Astronomy and Astrophysics}\ }\textbf {\bibinfo {volume} {36}},\ \bibinfo {pages} {599} (\bibinfo {year} {1998})}\BibitemShut {NoStop}%
\bibitem [{\citenamefont {Brandenberger}(1995)}]{brandenberger1995formation}%
  \BibitemOpen
  \bibfield  {author} {\bibinfo {author} {\bibfnamefont {R.~H.}\ \bibnamefont {Brandenberger}},\ }\href@noop {} {\bibfield  {journal} {\bibinfo  {journal} {arXiv preprint astro-ph/9508159}\ } (\bibinfo {year} {1995})}\BibitemShut {NoStop}%
\bibitem [{\citenamefont {Peacock}(2002)}]{peacock2002introduction}%
  \BibitemOpen
  \bibfield  {author} {\bibinfo {author} {\bibfnamefont {J.~A.}\ \bibnamefont {Peacock}},\ }\href@noop {} {\bibfield  {journal} {\bibinfo  {journal} {Electron--Positron Physics at the Z}\ ,\ \bibinfo {pages} {9}} (\bibinfo {year} {2002})}\BibitemShut {NoStop}%
\bibitem [{\citenamefont {Ma}\ and\ \citenamefont {Bertschinger}(1995{\natexlab{a}})}]{Ma:1995ey}%
  \BibitemOpen
  \bibfield  {author} {\bibinfo {author} {\bibfnamefont {C.-P.}\ \bibnamefont {Ma}}\ and\ \bibinfo {author} {\bibfnamefont {E.}~\bibnamefont {Bertschinger}},\ }\href {https://doi.org/10.1086/176550} {\bibfield  {journal} {\bibinfo  {journal} {Astrophysical Journal}\ }\textbf {\bibinfo {volume} {455}},\ \bibinfo {pages} {7} (\bibinfo {year} {1995}{\natexlab{a}})},\ \Eprint {https://arxiv.org/abs/astro-ph/9506072} {arXiv:astro-ph/9506072} \BibitemShut {NoStop}%
\bibitem [{\citenamefont {Dodelson}(2003{\natexlab{a}})}]{Dodelson:2003}%
  \BibitemOpen
  \bibfield  {author} {\bibinfo {author} {\bibfnamefont {S.}~\bibnamefont {Dodelson}},\ }\href@noop {} {\emph {\bibinfo {title} {Modern Cosmology}}}\ (\bibinfo  {publisher} {Academic Press},\ \bibinfo {year} {2003})\BibitemShut {NoStop}%
\bibitem [{\citenamefont {Ma}\ and\ \citenamefont {Bertschinger}(1995{\natexlab{b}})}]{Ma_Bertschinger_1995}%
  \BibitemOpen
  \bibfield  {author} {\bibinfo {author} {\bibfnamefont {C.-P.}\ \bibnamefont {Ma}}\ and\ \bibinfo {author} {\bibfnamefont {E.}~\bibnamefont {Bertschinger}},\ }\href {https://doi.org/10.1086/176550} {\bibfield  {journal} {\bibinfo  {journal} {The Astrophysical Journal}\ }\textbf {\bibinfo {volume} {455}},\ \bibinfo {pages} {7} (\bibinfo {year} {1995}{\natexlab{b}})},\ \Eprint {https://arxiv.org/abs/astro-ph/9506072} {astro-ph/9506072} \BibitemShut {NoStop}%
\bibitem [{\citenamefont {Dodelson}(2003{\natexlab{b}})}]{Dodelson_2003}%
  \BibitemOpen
  \bibfield  {author} {\bibinfo {author} {\bibfnamefont {S.}~\bibnamefont {Dodelson}},\ }\href@noop {} {\emph {\bibinfo {title} {Modern Cosmology}}}\ (\bibinfo  {publisher} {Academic Press},\ \bibinfo {address} {San Diego},\ \bibinfo {year} {2003})\BibitemShut {NoStop}%
\end{thebibliography}%

\end{document}